\documentclass[12pt]{article}
\usepackage{amssymb,amsmath}
\usepackage{graphicx} 
\usepackage{tikz}
\usepackage[T1]{fontenc}
\usepackage[utf8]{inputenc}
\usepackage[labelfont=bf]{caption}
\usepackage{color}
\numberwithin{equation}{section}
\newtheorem{theorem}{Theorem}[section]

\newtheorem{corollary}[theorem]{Corollary}
\newtheorem{conjecture}[theorem]{Conjecture}

\def\d{\partial}

\def\f{\frac}

\def\QEDclosed{\mbox{\rule[0pt]{1.3ex}{1.3ex}}} 

\def\QED{\QEDclosed} 
\def\endproof{\hspace*{\fill}~\QED\par\endtrivlist\unskip}
\newcommand{\eqa}{\begin{eqnarray}}
\newcommand{\eeqa}{\end{eqnarray}}
\newcommand{\beq}{\begin{equation}}
\newcommand{\eeq}{\end{equation}}

\begin{document}

\title{On integrable conservation laws}
\author{Alessandro Arsie$^{a)}$, Paolo Lorenzoni$^{b)}$, Antonio Moro$^{c)}$\\
\\
{\small $^{a)}$Department of Mathematics and Statistics}\\
{\small University of Toledo,}
{\small 2801 W. Bancroft St., 43606 Toledo, OH, USA}\\
{\small $^{b)}$Dipartimento di Matematica e Applicazioni}\\
{\small Universit\`a di Milano-Bicocca,}
{\small Via Roberto Cozzi 53, I-20125 Milano, Italy} \\
{\small $^{c)}$Department of Mathematics and Information Sciences}\\
{\small University of Northumbria at Newcastle,}\\
{\small Pandon Building, Camden Street NE2 1XE, Newcastle upon Tyne, UK}
}

\date{}

\maketitle

\begin{abstract}
We study normal forms of scalar integrable dispersive (non necessarily Hamiltonian) conservation laws via the Dubrovin-Zhang perturbative scheme. Our computations support the conjecture that such normal forms are parametrised by infinitely many arbitrary functions that can be identified with the coefficients of the quasilinear part of the equation. More in general, we conjecture that two scalar integrable evolutionary PDEs having the same quasilinear part are Miura equivalent. This conjecture is also consistent with the tensorial behaviour of these coefficients under general Miura transformations.
\end{abstract}

\section{Introduction}
The Dubrovin-Zhang perturbative approach is concerned with  the classification problem of evolutionary PDEs (in the scalar case) of the form
\beq\label{evosys}
u_t=X(u,u_x,...),\qquad i=1,\dots,n,
\eeq
where the functions $X(u,u_x,...)$ are differential polynomials in the jet variables $u_{x}$, $u_{xx}$, $\dots$. Introducing a rescaling of independent variables of the form $x\to\epsilon x$ and $t\to  \epsilon t$, the equation~(\ref{evosys}) takes the form
\beq\label{evosys2}
u_t=\sum_{k\geq -1} \epsilon^k F_{k+1}(u,u_x,...,u_{(k)},\dots),
\eeq
 where the functions $F_{k}$ are homogeneous differential polynomials of suitable degree and we adopt the notation $u_{(k)} := \partial_{x}^{k} u$. It is also assumed that
 \[
 F_{0}(u) \equiv 0.
 \]
 Hence, we will focus on a class of evolutionary PDEs of the form 
 \beq\label{evosys3}
u_t=\sum_{k\geq 0} \epsilon^k F_{k+1}(u,u_x,...,u_{(k)},\dots),
\eeq
where the r.h.s is a \emph{formal power series} in $\epsilon$ and it does not necessarily truncate.
Introducing a gradation such that  functions depending on the single variable $u$ have degree zero and monomials of the form     
 $u_{(k)}$ have degree $k$, it is straightforward to check that  the differential polynomial  $F_{k}$ has degree $k$. For instance, we have
\begin{eqnarray*}
F_{1}&=&V(u)u_x\\
F_{2}&=&A(u)u_{xx}+B(u)u_x^2.
\end{eqnarray*}
The Burgers equation
\beq
\label{burg}
u_{t} = u u_{x} + \epsilon u_{xx}
\eeq
and the Korteweg-de Vries (KdV) equations
\beq
\label{KdV}
u_{t} = u u_{x} + \epsilon^{2} u_{xxx}
\eeq
are two celebrated examples of exactly integrable nonlinear PDEs of the form~(\ref{evosys3}). As the r.h.s. of equation~(\ref{evosys3}) is allowed to be an infinite power series in $\epsilon$, the class of equations under study also includes non-evolutionary examples such as the Camassa-Holm equation~\cite{CH}
\beq\label{CH}
u_t-\epsilon^2u_{xxt}=-3uu_x+\epsilon^2(uu_{xxx}+2u_xu_{xx}).
\eeq
Indeed, the Camassa-Holm equation~(\ref{CH}) can be recast in the evolutionary form via the transformation
$$v=u-\epsilon^2u_{xx}=(1-\epsilon^2\d_x^2)u,$$
whose formal inverse is given by
$$u=(1-\epsilon^2\d_x^2)^{-1}v=1+\epsilon^2 v_{xx}+\epsilon^4 v_{xxxx}+\dots.$$
One of the main problem in the theory of integrable PDEs is to classify equations (or systems of equations) of the form~(\ref{evosys3}) up to equivalence under the so-called \emph{Miura transformations}
$$u\to\tilde{u}=M_0(u)+\sum_k\epsilon^k M_k(u,u_x,\dots),\qquad {\rm deg}\,M_k=k.$$
Hence, the classification problem of integrable equation of the form~(\ref{evosys3}) is reformulated in terms of a classification problem of equivalence classes of integrable equations with respect to Miura transformations. 
The Dubrovin-Zhang perturbative scheme aims at the reconstruction of higher order {\it integrable} corrections (both dispersive and dissipative ) starting from the quasilinear PDE of Hopf type (the dispersionless limit)
$$u_t=f(u)u_x.$$

Within this scheme, the various perturbative approaches developed so far mainly differ in the kind of additional structures possessed by the dispersionless limit and that are required to be preserved by the perturbation procedure.

Let us consider for instance the Hopf equation $u_t=uu_x$. It is clearly integrable as it possesses infinitely many symmetries  parametrised by an arbitrary function of one variable $g(u)$. The most general approach to the classification of integrable deformations of the Hopf equation is based of the request that all deformed symmetries $u_{\tau}=g(u)u_x+...$ commute with  the deformed Hopf equation $u_t=uu_x+...$~\cite{S,LZ}.
 
The classification of integrable conservation laws is based on the simple observation that the Hopf hierarchy consists of conservation laws on the form $u_t=\d_x\left(G(u)\right)$ and one may require that the deformation of the integrable hierarchy preserves the form of conservation law, i.e. $u_{t}=\d_x \left(G(u)+...\right)$. The general classification of scalar viscous conservation laws has been recently discussed in~\cite{ALM}.

A special class of conservation laws is given by Hamiltonian equations. These are equations of the form $u_{t} = \partial_{x} (G(u) + \dots)$ such that the deformed currents $G(u) + \dots$ can be written as variational derivatives w.r.t. the variable $u$, i.e.
$$ u_{t}=\d_x\f{\delta}{\delta u}\left(\int (h_0(u) + \dots )\,dx\right).$$
At the dispersionless level all equations of Hopf hierarchy are Hamiltonian w.r.t. the operator $\d_{x}$ that defines a Poisson bracket of hydrodynamic type~\cite{DN}.
This observation suggests to deform the Hamiltonians, requiring that they remain in involution w.r.t. the Poisson bracket. This approach has been first proposed and developed in~\cite{D}.

An alternative classification procedure relies on the observation that Hopf type equations possess a bi-Hamiltonian structure. 
This suggests to classify integrable deformations according to the existence of a deformed bi-Hamiltonian structure~\cite{DZ,L,DLZ,LZ,LZ2,LZ3}.

A common feature of these different approaches is that deformations are parametrised by 
  arbitrary functions. Clearly, the numbers of the functional parameters involved crucially depends on the problem
 at hand. In this paper, following \cite{ALM} we consider the case of scalar conservation laws
  extending the analysis to the case of dispersive conservation laws. Besides the undeniable relevance of
 conservation laws in physical applications, our focus is also motivated by the fact that, within the more general context of systems of PDEs of hydrodynamic type, the class of integrable diagonalizable equations \cite{Tsarev} coincide with the class of diagonalisable
  systems of conservation laws \cite{Sevennec}.
     
A key observation of the present work is that for scalar evolutionary PDEs the coefficients corresponding to the quasilinear terms have a tensorial nature. More precisely, given a PDE of the form
\[
u_t=X_1(u)u_x+\epsilon(X_2(u)u_{xx}+\dots)+\epsilon^2(X_3(u)u_{xxx}+\dots)+ \dots
\]
the coefficients $X_{1}(u)$, $X_{2}(u)$, etc. of the quasilinear terms \emph{are invariant} under Miura transformations of the form:  
\beq
u\to v=u+\sum_k\epsilon^k M_k(u_x,u_{xx},\dots)
\eeq  
 It is thus natural to expect that these coefficients play a crucial role in the classification problem. We also observe that the above transformations, which can be seen as perturbation of the identity, trivially preserve the dispersionless limit.

Based on results of the present paper combined with results already existing in the literature we formulate the following conjecture:

\begin{conjecture}\label{main}
Two scalar integrable evolutionary PDEs admitting the same quasilinear part are Miura equivalent.
\end{conjecture} 
 
According to this conjecture the number of free functional parameters appearing in deformations coincide with the number of independent functions in the quasilinear part of the equation.
 

\section{Tensorial coefficients}
In this Section we analyse in more detail the transformation properties of quasilinear terms in evolutionary equations of the form
\beq\label{scalar}
u_t=X(u,u_x,u_{xx},\dots)=X_1(u)u_x+\epsilon(X_2(u)u_{xx}+\dots)+\epsilon^2(X_3(u)u_{xxx}+\dots)+\mathcal{O}(\epsilon^3) 
\eeq 
under a Miura tranformation of the following type
\beq\label{Miura}
u\to v=M_0(u)+\sum_k\epsilon^k M_k(u_x,u_{xx},\dots).
\eeq
Observing that the vector field $X(u,u_x,u_{xx},\dots)$ in equation~(\ref{scalar}) transforms according to the rule
\beq
X(u,u_x,...)\to \tilde{X}(v,v_x,...)=\left(\f{\d v}{\d u}+\f{\d v}{\d u_x}\d_x+\f{\d v}{\d u_{xx}}\d_x^2+\dots\right)X_{u=u(v,v_x,\dots)}
\eeq
where  $u=u(v,v_x,\dots)$ is the inverse of Miura transformation \eqref{Miura}, we show that the coefficients of leading derivatives $X_{1}(u)$, $X_{2}(u)$, etc in \eqref{scalar} are not affected by the corrections to the leading part $M_{0}(u)$ of the Miura transformation~(\ref{Miura}). More precisely, these coefficients transform as tensors w.r.t the leading term of the transformation and in particular are invariant if such a leading term is the identity, i.e.
\beq\label{Miuraspecial}
u\to v=u+\sum_k\epsilon^k M_k(u_x,u_{xx},\dots).
\eeq 
We can prove the following
\begin{theorem}
\label{mainth}
Under the Miura transformation \eqref{Miura}, the coefficients $X_1(u),X_2(u),X_3(u),...$ of the quasilinear terms in the right hand side of \eqref{scalar}
 transform as
$$X_k(u)\to \tilde{X}_k(v)=X_k(u(v))$$
where $u(v)$ is the inverse of the dispersionless limit of \eqref{Miura}.
\end{theorem}

\noindent
\emph{Proof}. First of all we observe that quasilinear terms in the differential polynomial 
$$\tilde{X}(v,v_x,...)=\tilde{X}_1(v)v_x+\epsilon(\tilde{X}_2(v)v_{xx}+\dots)+\epsilon^2(\tilde{X}_3(v)v_{xxx}+\dots)+O(\epsilon^3)$$ are completely determined by quasilinear terms in the differential polynomial 
\begin{equation}\label{eqMiura3}\tilde{X}(u,u_x,...)=\left(\f{\d v}{\d u}+\f{\d v}{\d u_x}\d_x+\f{\d v}{\d u_{xx}}\d_x^2+\dots\right)X(u,u_x,...)\end{equation}
$$=\tilde{X}_1(u)u_x+\epsilon(\tilde{X}_2(u)u_{xx}+\dots)+\epsilon^2(\tilde{X}_3(u)u_{xxx}+\dots)+O(\epsilon^3).$$
Observing that  the inverse of \eqref{Miura} is of the form 
\begin{equation}
\label{inverseMiura}u=N_0(v)+\sum_{k\geq 1}\epsilon^k N_k(v, v_x, \dots)
\end{equation}
it can be easily proved by induction that
 term $\tilde{X}_k(v)v_{(k)}$ (i.e. the quasilinear term in $\tilde{X}(v,v_x,...)$ of degree $k$) is determined by quasilinear terms in $\tilde{X}(u,u_x,...)$ of degree less than or equal to $k$. Hence, in the following we will focus our analysis on quasilinear terms of
 $\tilde{X}(u,u_x,...)$ only. 
 Let us now write the Miura transformation \eqref{Miura} as
   \begin{equation}\label{eqMiura2}
   u\mapsto v=M_0(u)+\sum_{k\geq 1}\epsilon^k(a_k(u)u_{(k)}+R_k(u, u_x, \dots)),
   \end{equation}
  where the homogenous part of the $k-$th degree $M_k(u, u_x, \dots)$ has been decomposed into the quasilinear part and the remainder. We check by a direct calculation that quasilinear term of $\tilde{X}_m u_{(m)}$ of $\tilde{X}(u,u_x,...)$ computed by using the formula~\eqref{eqMiura3} are not affected by the reminder $R_k$.
   
Let us now write the vector field  $X(u, u_x, \dots)$ as 
$$X=\sum_{l\geq 1}\epsilon^{l-1}X_l(u)u_{(l)}+\mathrm{NQ}$$
where $\mathrm{NQ}$ denotes the non-quasilinear part of $X(u, u_x,\dots)$ and compute the transformed vector field~\eqref{eqMiura3}
  $$\tilde{X}(u, u_x, \dots)=\left(\frac{\partial F_0(u)}{\partial u}+\sum_{k\geq 1}\epsilon^k a_k(u)\partial_x^k+\mathcal{R}_1 \right)\left( \sum_{l\geq 1}\epsilon^{l-1}X_l(u)u_{(l)}+\mathrm{NQ}\right)$$
   where $\mathcal{R}_1$ accounts the terms produced by the remainders $R_k$ in \eqref{eqMiura2} and terms of the type $\sum_{k\geq1} \epsilon^k \frac{\partial a_k(u)}{\partial u}u_{(k)}$. 
Observing that the action of $\mathcal{R}_1$ on $X(u, u_x,\dots)$ always produces non-quasilinear terms we have
   \begin{equation}\label{transf2}\tilde{X}(u, u_x, \dots)=\sum_{l\geq 1}\epsilon^{l-1}X_l(u)\left(\frac{\partial F_0(u)}{\partial u}u_{(l)}+\sum_{k\geq 1}\epsilon^k a_k(u)u_{(k+l)}+\mathrm{NQ1} \right)+\mathrm{NQ2},\end{equation}
where $\mathrm{NQ1}$ denotes non-quasilinear terms produced by $\mathcal{R}_1$ and $\mathrm{NQ2}$ stays for remaining non quasilinear terms.
We can now evaluate explicitly quasilinear terms in~\eqref{transf2}.
Observing that
$$\d_x^l v=\f{\d F_0}{\d u}u_{(l)}+\sum_k\epsilon^k(a_k(u)u_{(k+l)})+\mathcal{R},$$
where $\mathcal{R}$ contain products of at least two derivatives of $u$, 
we have that the bracket in~\eqref{transf2}
$$\frac{\partial F_0(u)}{\partial u}u_{(l)}+\sum_{k\geq 1}\epsilon^k a_k(u)u_{(k+l)}+\mathrm{NQ1}$$ 
is equal to 
$$\partial_x^l v+\mathrm{NQ1}-\mathcal{R}$$ and therefore 
$$\tilde{X}(v, v_x, \dots)=\sum_{l\geq 1}\epsilon^{l-1}X_l(u)_{|u=u(v)}\partial_x^l v+\dots,$$
where the dots stands for non-quasilinear terms, and $u=u(v)$ is the inverse of the dispersionless part of the Miura transformation.  
\endproof

We have also the following

\begin{corollary}
If two evolutionary PDEs are Miura equivalent and have the same dispersionless limit then their quasilinear parts
 coincide.
\end{corollary}

Obviously, the converse statement is in general not true. However we conjecture that it is valid if one restricts
 to the class of \emph{integrable} equations (see Conjecture \ref{main}). 
 
\section{Scalar conservation laws}
This Section is devoted to the study of integrable scalar conservation laws of the form
\begin{equation}\label{Dcl}
u_t=\d_x\left[g(u)+\sum_{k=1}^{\infty}\epsilon^{k}\omega_{k}(u,u_x,...)\right],
\end{equation}
where $\omega_{k}$ are differential polynomials of degree $k$. 

For the sake of simplicity we will focus on the case $g(u)=u^2$. The general case can be treated in an analogous way. 
 
In virtue of Conjecture (\ref{main}), equivalence classes (with respect to the action of the Miura group) of integrable scalar equations
  are labelled by the independent coefficients of the quasilinear part. 
  Depending on the class of the equations considered, it might happen that only a subset of the coefficients of the quasilinear part are sufficient to determine all the others. The following analysis provides evidence of the fact that this is the case for {\em independent} coefficients of the quasilinear part in conservation laws of the form~(\ref{Dcl}). We will call them \emph{central invariants} by analogy with the central invariants introduced in~\cite{DLZ}.
    
 We follow the approach presented in~\cite{ALM} for viscous conservation laws, that is the case $\omega_{1} \neq 0$ in~(\ref{Dcl}), and extend it to dispersive conservation laws where only even powers in the formal parameter $\epsilon$ appear in~(\ref{Dcl}). The main steps of this approach can be summarised as follows:
\begin{enumerate}
\item Reduce of \eqref{Dcl} to its normal form 
\begin{equation}\label{clnew} 
u_t=\d_x\tilde\omega_{u^{2}}=\d_x\left[u^{2} +\epsilon a(u)u_x+\sum_{k>1} \epsilon^{k} \tilde{\omega}_{k}(u, u_{x}, \dots)\right]
\end{equation} 
where
$$
\frac{\partial \tilde \omega_{k}}{\d u_{x}}=0, \quad \forall k>1.
$$ 
This reduction is always possible and it is unique (see \cite{ALM}). 

\item Impose the \emph{integrability} condition,  i.e. the requirement that there exists a family of conservation laws 
\begin{equation}\label{DclSym}
u_\tau=\d_x\omega_{f}^{def}=\d_x\left[f(u)+\sum_{k=1}^{\infty}\epsilon^{k}f_{k}(u,u_x,...)\right]
\end{equation}
that commute with \eqref{Dcl}. We note that, as shown in \cite{AL,ALM}, this is equivalent to require that the 1-forms 
 $\omega_{u^{2}}$ and $\omega_{f(u)}$ are in involution w.r.t. the Poisson bracket
\beq
\label{ALbracket}
\{\alpha,\beta \} := \sum_{j} \partial_{x}^{j+1} \beta \frac{\partial \alpha}{\partial u_{(j)}} - \partial_{x}^{j+1} \alpha \frac{ \partial \beta}{\partial u_{(j)}}  =0.
\eeq     
In general one imposes the commutativity up to a fixed order in $\epsilon$ and one derives relations that express the terms $f_k$ appearing in \eqref{DclSym} as functions of the terms $a(u)$ and $\omega_k$ appearing in \eqref{clnew} and of the leading term $f(u)$ in \eqref{DclSym}. Depending on the structure of the equation \eqref{clnew} under consideration, there might be different constraints among $a(u)$ and the coefficients in $\omega_k$, as we will see below. The presence or absence of these constraints will single out the {\em independent} coefficients of the quasilinear part.
\end{enumerate}

\subsection{The viscous case}
Let us briefly review the case of a scalar conservation law with viscosity studied in \cite{ALM} and that corresponds to the assumption $a(u)\ne 0$ in \eqref{clnew}.
The procedure outlined above leads to the following
 \begin{theorem}\label{classth}
Up to $\mathcal{O}(\epsilon^{6})$, the quasilinear part of $\omega^{def}_{u^2}$
\begin{gather}
\label{normalform}
\begin{aligned}
 u^{2} + \epsilon a(u) u_{x} + \epsilon^{2}b_{1}(u) u_{xx} +\epsilon^{3}c_{1}(u) u_{xxx}+\epsilon^{4}d_{1}(u) u_{4x}+\epsilon^{5}e_{1}(u) u_{5x}+\mathcal{O}(\epsilon^6) \\
\end{aligned}
\end{gather}
is uniquely determined by $a(u)$. More precisely we have
$$b_1=(a^2/2!)',\,c_1=(a^3/3!)'',\,d_1=(a^4/4!)''',\,e_1=(a^5/5!)'''' \dots $$
\end{theorem} 
As a consequence, up to order $\mathcal{O}(\epsilon^{5})$, $a(u)$ is the only independent coefficient and it is named {\em viscous central invariant} in \cite{ALM}. Furthermore, the coefficients $f_k$ of the symmetries \eqref{DclSym} are also completely determined by $a(u)$ and by the leading term of the symmetries $f(u)$.

This result suggests the following 
\begin{conjecture}\label{qpvcl}
The quasilinear part of a viscous conservation law \eqref{clnew} is uniquely determined by $a(u)$.
\end{conjecture} 
Accordingly the main Conjecture in the case of scalar viscous conservation laws can be formulated as follows:
\begin{conjecture}\label{vcl}
Two integrable viscous conservation laws \eqref{clnew} admitting the same \emph{viscous central invariant} $a(u)$ are Miura equivalent.
\end{conjecture} 
Therefore, if this conjecture is true, for scalar viscous conservation laws \eqref{clnew} there exists only one independent coefficient in the quasilinear part of the equation. 

\subsection{Dispersive case} 
The case of dispersive conservation laws arises as a branching in the classification procedure that corresponds to the choice   $a(u)=0$.  The difference with the viscous case (see Theorem \ref{classth}) is remarkable as the classification suggests the existence of infinitely many free functional parameters.  Let us assume for simplicity that the coefficients in front of all odd powes of $\epsilon$ vanish (we will justify later this assumption). In this case the current in \eqref{clnew} reads as
\begin{gather}
\label{normalform}
\begin{aligned}
\omega^{def}_{u^2} &= u^{2} + \epsilon^{2}b_{1}(u) u_{xx} + \epsilon^{4} 
\left[c_{1}(u) u_{4x} + c_{2}(u) u_{xx}^2\right]+\\ 
&+ \epsilon^{6} \left[d_{1}(u) u_{6x} + d_{2}(u) u_{xx} u_{4x}+ d_{3}(u) u^2_{xxx}+ d_{4}(u) u^3_{xx} \right]+\\
&+ \epsilon^{8} \left[e_{1}(u) u_{8x} + e_{2}(u) u_{xx} u_{6x}+ e_{3}(u) u_{5x}u_{xxx}
+ e_{4}(u) u^2_{4x}+\right.\\
&\left.+e_5(u)u_{4x}u_{xx}^2+e_6(u)u^2_{xxx}u_{xx}+e_7(u)u_{xx}^4\right]+\dots,
\end{aligned}
\end{gather}
and the current in \eqref{DclSym} 
\begin{gather}
\label{normaldeformed}
\begin{aligned}
\omega^{def}_f&=f(u)+\epsilon^{2}\left[B_{1}(u)u_{xx}+ B_{2}(u)u_{x}^{2}\right]+\\ 
&+\epsilon^{4}\left[C_{1}(u)u_{4x}+C_{2}(u)u_{x}u_{xxx}+C_{3}(u)u_{xx}^{2}+C_{4}(u)u_{x}^{2}u_{xx}+C_{5}(u)u_{x}^{4}\right]\\
&+\epsilon^{6}\left[D_{1}(u)u_{6x}+D_{2}(u)u_{x}u_{5x}+D_{3}(u)u_{xx}u_{4x}+D_{4}(u)u_{x}^{2}u_{4x} 
+D_{5}(u)u_{xxx}^{2}+\right.\\
&\left.+D_{6}(u)u_{x}u_{xx} u_{xxx}+D_{7}(u)u_{x}^{3}u_{xxx}+D_{8}(u)u_{xx}^{3}+D_{9}(u)u_{xx}^{2}u^2_{x}+D_{10}(u)u_{xx}u^4_{x}+\right.\\
&\left.+D_{11}(u)u^6_{x}\right]+\dots
 \end{aligned}
\end{gather}
The integrability condition, that is the involutivity conditions on the associated $1$-forms 
$$\{\omega^{def}_{u^2},\omega^{def}_{f(u)}\}=0,\quad\forall\, f(u)$$ 
up to the order $\mathcal{O}(\epsilon^{12})$,  gives the following set of constraints\\

\noindent At order $\epsilon^{0}$ no conditions are enforced.\\

\noindent At order $\epsilon^{2}$,  $B_{1}$ and $B_{2}$ are expressed in terms of $b_{1}$ and $f$.\\

\noindent At order $\epsilon^{4}$, $C_{1}$, $C_{2}$, $C_{3}$, $C_4$ and $C_5$ are expressed in terms of $b_1,c_1,c_2$ and $f$. \\

\noindent At order $\epsilon^{6}$ the terms $D_{i}$, $i=1,..,11$ are given as functions of $b_1,c_1,c_2,d_1,d_2,d_3,d_4$ and $f$. \\

\noindent At order $\epsilon^{8}$ the terms  $E_{i}$, $i=1,..,22$ are expressed as functions of the small letters (coefficients of $\omega^{def}_{u^2}$) and $f$. Moreover there appear constraints that express $c_2$ in terms of $b_1$, $c_1$ and $d_1$
\begin{equation}\label{c2}
c_2=\f{1}{144}\f{1}{b_1^2}\left[117\left(\f{\d^2 b_1}{\d u^2}\right)b_1^3
-84b_1^2\left(\f{\d b_1}{\d u}\right)^2+670\left(\f{\d b_1}{\d u}\right)c_1
-330b_1^2\left(\f{\d c_1}{\d u}\right)+560b_1d_1-800c_1^2\right]
\end{equation}
and $d_3$ and $d_4$ in terms of $b_1$, $c_1$, $d_1$ and $d_2$.\\

\noindent  At order $\epsilon^{10}$ all the terms $F_{i}$, $i=1,..,42$ are expressed as functions of the small letters (coefficients of $\omega^{def}_{u^2}$) and $f$. Moreover there appears a constraint that gives $d_2$ in terms of $b_1$, $c_1$ and $d_1$. Furthermore the coefficients $e_2$, $e_4$, $e_5$, $e_7$, $f_4$ and $f_6$ also are determined in terms of the other coefficients of $\omega^{def}_{u^2}$.\\
 
\noindent At order $\epsilon^{12}$ all the coefficients $G_{i}$, $i=1,..,77$ are expressed as functions of the small letters (coefficients of $\omega^{def}_{u^2}$) and $f$. Moreover there appear constraints that give $e_3,e_6$,  part of the $f_i's$ and part of $g_i's$ in terms of the remaining coefficients of $\omega^{def}_{u^2}$, namely those coefficients that are still free.  \\

\noindent Summarizing, the coefficients
 $c_2,d_2,d_3,d_4,e_2,e_3,e_4,e_5,e_6,e_7$ of  $\omega_{u^2}^{def}$ are uniquely determined in terms of $b_1,c_1,d_1,e_1$ and eventually of higher order small letters.\\
  
Above results are summarised in the following
\begin{theorem}\label{classnew}
Up to $\mathcal{O}(\epsilon^{12})$, all the coefficients of the quasilinear part of
 \eqref{normalform} are independent. 
\end{theorem}

For the sake of simplicity, we have imposed from the very beginning that in the case of $a(u)=0$ only even powers
 of $\epsilon$ are present. However one can check that this assumption is not restrictive and it follows directly from computations, at least up to the sixth order.

Above results lead to the following
\begin{conjecture}\label{qpvcl}
The quasilinear part of an integrable dispersive conservation law contains only even powers of $\epsilon$
 and all the coefficients of the quasilinear part are independent. 
\end{conjecture} 

\subsection{Are all dispersive conservation laws Hamiltonian?} 
In \cite{D} it was proved that Hamiltonian conservation laws can be reduced to the form
\beq
\label{ham}
u_t=\d_x\left[\f{u^2}{2}+\frac{\epsilon^2}{24}\left(2cu_{xx}+c'u_x^2\right)+
\epsilon^4\left(2pu_{(4)}+4p'u_xu_{xxx}+3p'u_{xx}^2+2p''u_x^2u_{xx}\right)+\mathcal{O}(\epsilon^6)\right]
\eeq
where $p(u)$ and $c(u)$ are arbitrary functions. Equation~(\ref{ham}) is brought to the normal form \eqref{clnew}, up to $\mathcal{O}(\epsilon^6)$,  
\beq
\label{normham}
u_t=\d_x\left[\f{u^2}{2}+\frac{\epsilon^2}{12} c(u) u_{xx}+
\epsilon^4\left(2p(u) u_{(4)} +\frac{4 c'(u)^{2} + 3 c c''}{1152} u_{xx}^2 \right)+\mathcal{O}(\epsilon^6)\right]
\eeq
by means of the Miura transformation
$$\tilde{u}=u+ \epsilon \d_x\left( \epsilon \alpha(u) u_{x} + \epsilon^{2} \left(\beta_0(u)u_{xxx}+\beta_{1}(u)u_xu_{xx}+\beta_2(u)u_x^3\right) + \mathcal{O}(\epsilon^{4}) \right),$$
where the coefficients $\alpha(u), \beta_0(u), \beta_1(u)$ and $\beta_2(u)$ are given by
\begin{align*}
&\alpha(u) = - \frac{c'(u)}{24} \\
&\beta_{0}(u) = - p'(u) + \frac{c'(u)^{2} - c(u) c''(u)}{384} \\
&\beta_{1} (u) = -\frac{p''(u)}{2} + \frac{5 c'(u) c''(u) - 6 c(u) c'''(u)}{1152} \\
&\beta_{2}(u) = \frac{3 c''(u)^{2} + 2 c'(u) c'''(u) - 4 c (u) c^{(IV)}(u)}{3456}.
\end{align*}
Hence, the coefficients of the normal form~(\ref{normalform}) can be uniquely expressed in terms of the invariants $c(u)$ and $p(u)$. In particular, we have
\begin{align*}
&b_{1}(u) =  \frac{c(u)}{12} \\
&c_{1}(u) = 2  p(u) \\
&c_{2}(u) = \frac{4 c'(u)^{2} + 3 c(u) c''(u)}{1152}
\end{align*}
If, for instance, $c (u)$ is a non vanishing constant then the equation reads as
\beq
u_t=\d_x\left[\f{u^2}{2}+\frac{\epsilon^2}{24}\left(2cu_{xx}\right)+
\epsilon^4\left(2pu_{(4)}\right)+\mathcal{O}(\epsilon^6)\right]
\eeq 
Then, the relation~\eqref{c2} results into a constraint on the coefficient $d_{1}$ that is consequently no longer free being uniquely determined in terms of the two functional parameters $c(u)$ and $p(u)$, i.e.
\begin{eqnarray*} 
d_{1}(u) = \frac{1}{7} \frac{p(u)}{c(u)} \left[ 480 p(u) - \frac{67}{4} c'(u) \right] + \frac{1}{112} c(u) \left [ 11 p'(u) + \frac{13}{ 720} c'(u)^{2} - \frac{7}{960} c(u) c''(u) \right].
\end{eqnarray*}
Comparing this result with the one presented in Theorem \ref{classnew}, it follows that integrable hierarchies of dispersive conservation laws in general are not  Hamiltonian with respect to the Poisson operator $\d_x$ or with respect to a Poisson operator related to $\d_x$ by a Miura transformation. This fact does not exclude \emph{a priori} the possibility that a single equation of the hierarchy is Hamiltonian. An example of this type is the Sawada-Kotera hierarchy, where the SK equation is Hamiltonian but none of its symmetries can be obtained as a Hamiltonian deformation. 

\section{Examples}
The classification procedure discussed above turns out to be consistent with alternative definitions of integrability (e.g. S-integrability, existence of a bi-Hamiltonian structure, Backlund transformations) and reproduces a number of relevant examples known in the literature. Given the general integrable conservation law in the form
\begin{gather}
\begin{aligned}
u_t&=\d_x\left[u^{2} + \epsilon^{2}b_{1}(u) u_{xx} + \epsilon^{4} 
\left(c_{1}(u) u_{4x} +...\right)+\right.\\
&\left. + \epsilon^{6} \left(d_{1}(u) u_{6x} +...\right)+ \epsilon^{8} 
\left(e_{1}(u) u_{8x} +...\right)+...\right].
\end{aligned}
\end{gather}
for particular choices of the free functional parameters we easily recover a few examples well known in the literature. The list below is not meant to be complete.

\noindent {\bf KdV equation}. In the particular case of constant central invariants we could reproduce two important examples. The choice
\[
b_1=\textup{const.} \qquad  c_1(u)=d_1(u)=\cdots=0
\]
gives the celebrated Korteweg-de Vries equation.  \\

\noindent {\bf Hodge KdV equation}. The following example of constant central invariants
\[
b_1=\f{|B_2|}{2} \qquad c_1=\f{|B_4|}{4} \qquad d_1=\f{|B_6|}{6}...
\]
where $B_2=\f{1}{6},B_4=-\f{1}{30},...$ are Bernoulli numbers corresponds to the intriguing example of the Hodge KdV equation  that appeared in study of Gromow-Witten invariants in Topological Field Theory~\cite{B,Bu}. \\

\noindent {\bf Camassa-Holm and Degasperis-Procesi equations.}
In the case of linear central invariants we have the Camassa-Holm equation 
\[
u_t-\epsilon^2u_{xxt}=-3uu_x+\epsilon^2(uu_{xxx}+2u_xu_{xx})
\]
and Degasperis-Procesi equation \cite{DP}
\[
u_t-\epsilon^2u_{xxt}=-4uu_x+\epsilon^2(uu_{xxx}+3u_xu_{xx}).
\]
Notice that these two equations do not appear in the evolutionary form, but can be
brought to the evolutionary form via formal inversion of the operator $1-\partial^2_{x}$. They can also be reduced to the same dispersionless limit via a rescaling of the
dependent variable $u$. In both cases the  central invariants have the form 
\[
b_1(u)=c_1(u)=d_1(u)=\cdots=cu
\]
but the value of constant $c$ is different and thus as a consequence of Conjecture~\ref{main} they are not expected to be Miura equivalent.
\newline
\newline
\noindent We conclude this section presenting a list of examples of integrable equations sharing one and the same quasilinear part and that are known to be Miura equivalent.\\

\noindent {\bf KdV vs mKdV equation} 

\noindent
Let us consider the Korteweg-de Vries equation
\begin{equation}
\label{exKdV}
u_{t} = 2 u u_{x} + \epsilon^{2} u_{xxx}
\end{equation}
and the modified KdV equation in the form
\begin{equation}
\label{exmKdV}
v_{t} = - 3 v^{2} v _{x}  + \epsilon^{2} v_{xxx}.
\end{equation}
We observe that, introducing the transformation of the dependent variable $v = \pm \sqrt{- 2w/3}$, the equation~(\ref{exmKdV}) takes the form
\begin{equation}
\label{exmKdV2}
w_{t}  = 2 w w_{x} + \epsilon^{2} \left(w_{xxx} - \frac{3}{2w} w_{x}  w_{xx} + \frac{3}{4 w^{2}} w_{x}^{3}  \right),
\end{equation}
whose quasilinear part coincides with KdV equation~(\ref{exKdV}). Hence the Conjecture~\ref{main} is consistent with the very well known fact that there exists the Miura transformation mapping the equation~(\ref{exKdV}) into~(\ref{exmKdV}) (see e.g.\cite{A}), that is explicitly given by
\[
u = - \frac{3}{2} \left(v^{2} + \epsilon \sqrt{2} v_{x} \right).
\]

\noindent {\bf KdV vs Gardner equation} \\
The Gardner equation
\begin{equation}
\label{exGardner}
v_{t}  = \partial_{x} \left(v^{2} - \alpha v^{3} + \epsilon^{2} v_{xx} \right)
\end{equation}
is completely integrable and it  is known to be Miura equivalent to the KdV equation~(\ref{exKdV}) via the following invertible transformation~\cite{MGK}
\[
u = v - \frac{3}{2} \alpha v^{2}  - \frac{3}{2} \sqrt{2 \alpha} \epsilon v_{x}.
\]
We note that this transformation clearly does not preserve the dispersionless limit. In order to verify the consistency the Conjecture~\ref{main} with the above classical result it is first necessary to reduce the dispersionless part of Gardner's equation~(\ref{exGardner}) to the Hopf equation.
This is done by introducing the variable $w$  such that $v = (1\pm \sqrt{1 -6 \alpha w})/(3\alpha)$, such that the equation~(\ref{exGardner}) takes the form
\[
w_{t} = 2 w w_{x} + \epsilon^{2} \left(w_{xxx} + \frac{9 \alpha}{1 - 6 \alpha w} w_{x} w_{xx} + \frac{27 \alpha^{2}}{(1-6 \alpha w)^{2}} w_{x}^{3} \right).
\]
We immediately see that the above equation and the KdV equation~(\ref{exKdV}) share one and the same quasilinear part. \\

\noindent {\bf SK vs KK equation} \\
\noindent  A direct comparison of Sawada-Kotera and
$$u_t=\d_x\left[\f{5}{3}u^3+\epsilon^2(5uu_{xx})+\epsilon^4(u_{xxxx})\right]$$
Kaup-Kuperschmidt equations
$$u_t=\d_x\left[\f{5}{3}u^3+\epsilon^2\left(5uu_{xx}+\f{15}{4}u_x^2\right)+\epsilon^4(u_{xxxx})\right]$$
clearly show that these two equation possess the same quasilinear part and as a consequence of Conjecture~(\ref{main}) they are expected to be Miura equivalent. Indeed, such a Miura transformation exists and was discovered by Fordy and Gibbons~(\cite{FG}).\\

\section{Concluding remarks}

In this paper we focussed our attention on the study of equivalence classes of integrable dispersive conservation laws with respect to Miura transformations. The analysis of transformation properties of quasilinear terms in the deformation and the results from the perturbative approach  suggest that equations sharing one and the same quasilinear part  are also Miura equivalent. This seems to be a general principle. 
  
The table~(\ref{tab_results}) aims at providing a summary of currently known classification results of scalar integrable PDEs. We have also included the case of bi-Hamiltonian structures although
 in this case the invariant parameter is not directly related to the quasilinear part of the corresponding equations.
 
\begin{table}[h]
\begin{center}
\begin{tabular}{|l|c|l|}

    \hline
	{\bf Type of deformations} & {\bf Numbers of invariants} &
	  {\bf References}    \\ 
	\hline
General viscous deformations  & $2$ (conjecture) & \cite{LZ}  \\
\hline
General dispersive deformations  &  & \cite{S}  \\
\hline
Viscous conservation laws  & $1$ (conjecture) & \cite{ALM}  \\
\hline
Dispersive conservation laws  & $\infty$ (conjecture) & present paper  \\
	\hline
Hamiltonian conservation laws  & $\infty$ (conjecture)  & \cite{D,Dtalk}  \\
\hline
Bi-Hamiltonian structures & $1$ (proved in \cite{LZ3}) &  \cite{DZ,L,LZ2,LZ3,AL2}  \\
\hline
\end{tabular}
\end{center}
\caption{\footnotesize Summary of known and conjectured results in the scalar case.}
\label{tab_results}
\end{table}

A part from the bi-Hamiltonian case where results at any order are already available, other results have been proved so far only up to a certain order in the deformation parameter $\epsilon$. The number of independent functions parametrising the quasilinear part depends on the type of deformations. In the case of dispersive conservation laws apparently the coefficients of the quasilinear part can  be arbitrarily chosen. A similar freedom has been observed for Hamiltonian conservation laws \cite{Dtalk}. Notice that Miura transformations involved in this case are canonical and the comparison with dispersive conservation laws is not immediate (as explained in Section 3.3).   
 Viscous deformations turns out to be much more rigid being parametrised by a single function of one variable  for viscous conservation laws~\cite{ALM} and by two functions for general viscous equations~\cite{LZ}. 
   
We point out  that although there is a certain flexibility in the choice of the functional parameters that characterise the deformation, it is more convenient to choose  invariant parameters as they allow a direct comparison between different equations not necessarily brought to their normal form.    
   
We conclude mentioning that  the case of integrable systems of conservation laws  of the form
\begin{equation*}
u^i_t=\partial_x\left[f^i({\bf u})+\epsilon(A^i_j({\bf u})u^j_{x})+\epsilon^2(B^i_j({\bf u})u^j_{xx}+C^i_{jk}({\bf u})u^j_xu^k_x)+\mathcal{O}(\epsilon^3)\right],\qquad i=1,\dots,n.
\end{equation*}  
 is completely open. For instance, even the generalization of the notion of normal form is not straightforward. Moreover it is not difficult to check that the coefficients of the quasilinear part do not transform as tensors under a general Miura transformation.  
 We plan to tackle this case in a future publication.

\section*{Ackowledgements} 
We thank Gennady El for useful discussions on the Gardner equation. 
This research has been partially supported by the London Mathematical Society  Visitors Grant (Scheme 2) Ref.No. 21226.
The research of AA is partially supported by the Faculty Development Funds of the College of Natural Sciences and Mathematics, University of Toledo.
PL is partially supported by the Italian MIUR Research Project \emph{Teorie geometriche e analitiche dei sistemi Hamiltoniani in dimensioni finite e infinite}.


\begin{thebibliography}{99}

\bibitem{A} M.J. Ablowitz and H. Segur, Solitons and the inverse scattering transform, SIAM, Philadelphia, 1981.

\bibitem{AL2} 
A. Arsie, P. Lorenzoni, \emph{On bi-Hamiltonian deformations of exact pencils of hydrodynamic type}, J. Phys. A 44 (2011), no. 22,

\bibitem{ALM} 
A. Arsie, P. Lorenzoni, A. Moro, \emph{Integrable viscous conservation laws},  {\tt arXiv:1301.0950} (2013).

\bibitem{AL} 
A. Arsie and P. Lorenzoni,  \emph{Poisson bracket on 1-forms and evolutionary PDEs}, Journal of Physics A, Mathematical and Theoretical, (2012). 

\bibitem{BGI} V.A. Baikov, R.K Gazizov and N.Kh. Ibragimov, {\emph Approximate symmetries and formal
linearization}, J. Appl. Mech. Tech. Phys. 30, no. 2 (1989) 204-212.

\bibitem{B} A. Brini, Open topological strings and integrable hierarchies, remodeling the A-model, Comm. Math. Phys., {\bf 312}(3), 735-780 (2012).

\bibitem{Bu}  A. Buryak, H. Posthuma, S. Shadrin, On deformations of quasi-Miura transformations and the Dubrovin-Zhang bracket, J. Geom. Phys. {\bf 62}(7), 1639-1651 (2012).

\bibitem{CH} R. Camassa and D. Holm, An integrable shallow water wave equation with peaked
solitons, Physical Review Letters, {\bf 71}(11), 1661-1664 (1993).

\bibitem{DP}  A. Degasperis and M. Procesi, ``Asymptotic integrability", in A. Degasperis, G. Gaeta, Symmetry and Perturbation Theory (Rome, 1998), River Edge, NJ: World Scientific, pp. 23-37 (1999).

\bibitem{DN} B.A. Dubrovin, S.P. Novikov,
 \emph{On Hamiltonian brackets of hydrodynamic type},
 Soviet Math. Dokl. {\bf 279:2} (1984) 294--297.

\bibitem{D} B. Dubrovin, \emph{Hamiltonian peturbations of hyperbolic systems of conservation laws II}, Comm. Math. Phys. Volume {\bf 267}, Number 1, 117-139, (2006).

\bibitem{DZ} B. Dubrovin, Y. Zhang, \emph{Normal forms of integrable
PDEs, Frobenius manifolds and Gromov-Witten invariants},   math.DG/0108160.

\bibitem{DLZ} B. Dubrovin, S.-Q.Liu, Y.Zhang, \emph{On hamiltonian perturbations of hyperbolic
systems of conservation laws, I: quasitriviality of bihamiltonian perturbations},
 Comm. Pure and Appl. Math. {\bf 59}, 559--615 (2006);
 
\bibitem{Dtalk}
B. Dubrovin, \emph{Integrable systems in small dispersion expansion}, talk given at the conference
  Physics and Mathematics of Nonlinear Phenomena 2013 (Gallipoli).

\bibitem{FG} A. Fordy and J. Gibbons, Some remarkable nonlinear transformations, Phys. Lett. A, {\bf 75}(5), 325 (1980).

\bibitem{LZ}
S-Q. Liu, Y. Zhang \emph{On Quasitriviality and Integrability of a Class of Scalar
Evolutionary PDEs} J. Geom. Phys. {\bf 57}, 101--119, (2006).

\bibitem{LZ2}
S-Q. Liu, Y. Zhang \emph{Deformations of semisimple bihamiltonian
structures of hydrodynamic type}, J. Geom. Phys. {\bf 54}(4), 427--453, (2005).

\bibitem{LZ3}
S-Q. Liu, Y. Zhang \emph{Bihamiltonian Cohomologies and Integrable Hierarchies I: A Special Case}
Communications in Mathematical Physics (2013) Volume 324, {\bf 3}, pp 897--935

\bibitem{L} P. Lorenzoni, \emph{Deformations of bi-Hamiltonian structures
of hydrodynamic type}, J. Geom. Phys. {\bf 44} (2002) 331--375.

\bibitem{MGK} R. M. Miura, C. S. Gardner, and M. D. Kruskal,
Korteweg-de Vries equation and generalizations. II. Existence of conservation laws and constants of motion, J. Math. Phys.
{\bf 9}, 1204 (1968).

\bibitem{Sevennec}
B. Sevennec, \emph{Géométrie des systèmes hyperboliques de lois de conservation}, Mémoires de la Société Mathématique de France, {\bf 56}, pag 1--125 (1994).

\bibitem{S} I.A.B.Strachan, \emph{Deformations of the Monge/Riemann hierarchy and approximately integrable systems},
 J. Math. Phys. {\bf 44} (2003) 251--262.
 
\bibitem{Tsarev} S.P. Tsarev, \emph{Geometry of Hamiltonian systems of
hydrodynamic type. Generalized hodograph method}, Izvestija AN USSR
Math. {\bf 54}, no. 5 (1990)  1048-1068. 
 
 
\end{thebibliography}
\end{document}